\documentclass[AMA,STIX1COL]{WileyNJD-v2}

\usepackage{amsmath}
\usepackage{subfig}

\usepackage{graphicx}
\usepackage{xspace}
\newcommand{\ichunk}{\emph{iCh}\xspace}
\newcommand{\spmv}{\texttt{spmv}\xspace}

\newcommand{\stealing}{\emph{stealing}\xspace}
\newcommand{\dynamic}{\emph{dynamic}\xspace}
\newcommand{\guided}{\emph{guided}\xspace}
\newcommand{\taskloop}{\emph{taskloop}\xspace}
\newcommand{\binlpt}{\emph{binlpt}\xspace}

\newcommand{\chunk}{\emph{chunk size}\xspace}

\newcommand{\synth}{\texttt{synth}\xspace}

\newcommand{\bfs}{\texttt{BF}\xspace}

\newcommand{\lava}{\texttt{LavaMD}\xspace}

\newcommand{\kmeans}{\texttt{Kmeans}\xspace}

\usepackage{sparklines}

\articletype{Article Type}%

\received{}
\revised{}
\accepted{}
        
\begin{document}

\title{An Adaptive Self-Scheduling Loop Scheduler}

\author[1]{Joshua Dennis Booth*}
\author[1]{Phillip Allen Lane}

\authormark{Joshua Dennis Booth \textsc{et al}}

\address[1]{\orgdiv{Department of Computer Science}, \orgname{The University of Alabama in Huntsville}, \orgaddress{\state{Alabama}, \country{USA}}}

\corres{*Corresponding \email{joshua.booth@uah.edu}}

\presentaddress{301 Sparkman Dr NW, \\ Huntsville AL 35899}

\abstract[Summary]{Many shared-memory parallel irregular applications, such as sparse linear algebra and graph algorithms, depend on efficient loop scheduling (LS) in a fork-join manner despite that the work per loop iteration can greatly vary depending on the application and the input. Because of the importance of LS, many different methods (e.g., workload-aware self-scheduling) and parameters (e.g., chunk size) have been explored to achieve reasonable performance, and many of these methods require expert prior knowledge about the application and input before runtime.  
This work proposes a new LS method that requires little to no expert knowledge to achieve speedups close to those of tuned LS methods by self-managing chunk size based on a heuristic of throughput and using work-stealing to recover from workload imbalances.
This method, named \ichunk, is implemented into libgomp for testing.
It is evaluated against OpenMP's guided, dynamic, and taskloop methods and is evaluated against BinLPT and generic work-stealing on an array of applications that includes: a synthetic benchmark, breadth-first search, K-Means, the molecular dynamics code LavaMD, and sparse matrix-vector multiplication.
On a 28 thread Intel system, \ichunk is the only method to always be one of the top three LS methods.
On average across all applications, \ichunk is within $5.4\%$  of the best method and is even able to outperform other LS methods for breadth-first search and K-Means. }

\keywords{irregular applications, loop scheduling, performance evaluation, OpenMP}

\maketitle

\section{Introduction}
\label{sec:intro}
Large multicore systems dominate parallel computing platforms.
One common technique to improve the performance of an application on these systems is loop-level parallelism, and the popularity of this loop-level parallelism is seen in the number of built-in parallel-for (\texttt{par\_for}) implementations in various languages and packages~\cite{omp,cilk}.
Therefore, the scalability of these applications depends on scheduling loop iterations among threads so that the workload, i.e., the amount of computations and memory accesses, is balanced. 
This problem is commonly referred to as the loop scheduling (LS) problem. 
When the workload is uniformly distributed among the loop iterations, the issue of scheduling iterations becomes fairly trivial.
However, irregular applications, which are common in high-performance computing (HPC) (e.g., sparse linear algebra and graph algorithms)~\cite{roofline,bcnuma,csr,Kabir2014}, do not have uniform workload distributions, and thus the assignment of iterations to cores or threads becomes extremely difficult.
In fact, distributing iterations with a nonuniform workload is an NP-hard problem~\cite{nphard}.
The common solution to this problem is self-scheduling, i.e., the threads and cores determine the task size or the number of iterations themselves rather than by the operating system or a global control unit.
Within this solution exists numerous methods and heuristics~\cite{factoring,binlpt,history,omptask1,omptask2} requiring fine-tuning of parameters and expert knowledge. 
The scalability of even a simple \texttt{par\_for} on nonuniformly distributed workloads depends on selecting the best self-scheduling method and fine-tuning of parameters, such as chunk size. 
If the selection and tuning are not done, the performance may be magnitudes different from the tuned solutions or may not speed up performance at all. 

This work presents a new self-scheduling method that achieves good scalable performance across a wide range of applications with various loop workload distributions, and thus, this work removes the importance of self-scheduling method selection with minimal fine-tuning.  
In general, self-scheduling methods can be classified as: \emph{on-demand} (i.e., those that assign tasks at runtime to improve imbalance), \emph{chunk size tuning} (i.e., those that break tasks into chunks that will minimize scheduling overheads), or \emph{workload aware} (i.e., those that use the known workload distribution to assign tasks to threads).
Methods in each classification have unique benefits related to the classification that matches different applications and loop workloads, but none work equally well on all applications and workloads.
The self-scheduling method in this work bridges the shortcomings of methods in each of these classifications by combining design techniques from the first two to produce one self-scheduling method that does well for an array of different applications and does not need to be explicitly informed of the workload distribution like workload aware. 
As such, this method saves time in both implementations (i.e., not having to precalculate workload distribution for the scheduling method) and tuning (i.e., other scheduling methods do not need to be considered).

In order to achieve this flexibility, this new method uses distributed work queues and work-stealing~\cite{cilk,kappi} from \emph{on-demand} methods to balance workloads, uses a newly developed adaptive \emph{chunk size tuning} to reduce scheduling overheads of both per-chunk assignment and locks while work-stealing, and uses a new inexpensive calculation of the distribution of iteration throughput during runtime to update the per-thread chunk size and inform stealing threads about local queues.  
As this method depends on automatically adapting a per-thread chunk size for the irregular application, we name the method \ichunk (\emph{i}rregular \emph{Ch}unk).
This work is the first to use this form of adaptive chunk size, and it is the first to apply this type of technique to improve work stealing based self-scheduling.

The rest of this paper is organized as follows.  
A brief introduction to LS is given in Section 2, along with an introduction to workload distribution patterns found in irregular applications.
Section 3 provides an overview of the algorithm and mechanics used by \ichunk.
A review of related work for LS and their connection to \ichunk is provided in Section 4.
Section 5 provides the experimental setup for the evaluation of \ichunk, while Section 6 provides the results.
Concluding thoughts and overview are given in Section 7.

\section{Background}
\label{sec:background}
This section describes the background related to loop scheduling (i.e., how the problem is defined and common methods), and irregular applications (i.e., the characteristics of irregular applications that make them difficult to schedule and insight that motivated \ichunk).

\subsection{Loop Scheduling}
The LS problem requires scheduling a set of $n$ independent iterations, $x_{i}$ where $i \in \{1, \dots , n\}$, onto $p$ threads, $t_{j}$ where $j \in \{1, \dots , p \}$, in a manner that minimizes the total parallel execution time.
Though this problem is NP-hard~\cite{nphard}, the importance of the problem has inspired numerous self-scheduling methods built around heuristics, theoretical insight, and anecdotal evidence. 
Most heuristics and anecdotal evidence make some assumptions about the workload distribution.
For example, many methods assume that the iteration workload does not vary much or assume workloads are pulled from a normal distribution~\cite{equal}.
Additionally, they focus on improving overall load balance while minimizing overheads related to utilizing queues or bookkeeping variables that estimate balance or chunk size.
Many self-scheduling methods try to keep the method itself as simple as possible to not increase overheads, and will normally chunk loop iterations linearly (i.e., loop iteration $x_{i+1}$ will usually be assigned to the same thread as loop iteration $x_i$) in order to aid memory reuse and locality. 
Most of the self-scheduling methods built around theoretical insight stem from work-stealing.
In work-stealing, each thread works on its portion of the loop iterations, and when a thread runs out of work it randomly steals one-half of the remaining tasks from some other thread (called the victim) that has iterations remaining.
Work-stealing is shown to be a 2-approximation\cite{nphard} of the optimum solution and is very popular in task-based systems like CILK~\cite{cilk}.
Despite being popular, many self-scheduling systems do not rely on work-stealing because of issues related to the overheads in stealing and preserving locality in memory accesses~\cite{kappi}. 
However, many of the self-scheduling methods being implemented in the KMP (i.e., a popular OpenMP library that many compilers utilize, such as Clang and Intel)
\footnote{https://github.com/llvm-mirror/openmp/blob/master/runtime/src/kmp.h} OpenMP library of LLVM~\footnote{https://llvm.org/} have reserved flags for possible work-stealing implementations.
In more detail, the three approaches (\emph{on-demand}, \emph{chunk size tuning}, and \emph{workload aware}) from the previous section can be further classified in the following manner.
\begin{itemize}
\item 
\emph{Pure Dynamic Self-Scheduling} and \emph{Chunk Self-Scheduling} assign iterations to threads from a central, shared queue in unit-sized chunks that do not change during execution~\cite{chunk}.
A small chunk size allows for better load balance as iterations are assigned at a finer granularity but comes with a high assignment overhead.
A larger chunk size assigns iterations with a more coarse granularity leading to worse load balancing but comes with a much lower assignment overhead.
Most of these methods are workload-unaware, but expert knowledge of the application, input, and system can help tune the chunk sizes, thus making it workload-aware.
Additionally, when the chunk size is fine-tuned, the application can achieve near-optimal load balancing. 
An example of this is the dynamic scheduling already implemented into OpenMP (\dynamic) and is commonly used in irregular applications~\cite{bcnuma,Kabir2014}.
\item
\emph{Guided Self-Scheduling} assigns chunks to threads from a central queue, much like \emph{Chunk Self-Scheduling}, but dynamically changes chunk size at runtime.
The size change is normally only based on the remaining number of iterations and threads making the method semi-workload-learning, which is sometimes called the Load Imbalance Amortization Principle~\cite{guided}. 
\item
\emph{Factoring Self-Scheduling} determines the number of iterations to be assigned not via a chunk size, but by a fraction of the number of remaining iterations.
Both \emph{Guided Self-Scheduling} and \emph{Factoring Self-Scheduling} work well when loop iteration workloads are normally or uniformly randomly distributed, and it has an almost averaging effect of workloads in these situations (i.e., they can use the amortization principle). 
An example of a commonly used method related to these last two is \guided in OpenMP.
\item
\emph{Workload-Aware Self-Scheduling} uses expert information, i.e., known information about the workload distribution that must be precalculated, to distribute the iterations among threads~\cite{binlpt}.  
A second-level method, such as work stealing, is commonly used to deal with variations caused by the system, such as frequency scaling and memory access latency.
\item
\emph{History-Based Self-Scheduling} tries to build up information about the workload during execution~\cite{history}.
In practice, most require a nested-loop structure, where the outermost loop serially executes the parallel innermost loop.
During the first few iterations of the outermost loop, a history is built that helps distribute the parallel iterations.
However, if the time does not exist to train or the workload changes for the innermost loop, then the method may have difficulties achieving good speedups.
\end{itemize}

\subsection{Irregular Applications}
Many common HPC applications have subsections of code that make irregular accesses to memory in parallel, such as methods that deal with sparse matrices or graph algorithms.
These types of applications are commonly called irregular applications or Type-I applications~\cite{hpcg}.
Some common examples include graph mining and web crawls.
However, these applications require a great deal of tuning based on both the computer system and algorithm input to perform optimally. 
Many different programming models are used to implement these kernels, but one of the most common is a fork-join model.
Additionally, many of these kernels are memory-bound even when optimally programmed~\cite{roofline}.
This means that many memory requests are already waiting to be fulfilled, and additional requests will have high latency on an already busy memory system. 
In order to explore this, we will look at two important irregular applications: Breadth-First Search and Sparse Matrix-Vector Multiplication. \\

\noindent \textbf{Breadth-First Search.}
The breadth-first search application is commonly used inside many graph-analytic metrics, such as building spanning trees or calculating betweenness of centrality.
Additionally, many search-space problems that are solved using branching-and-bounding will explore the tree in a breadth-first search manner, and many scheduling algorithms that rely on task-levels for thread assignment and synchronization points explore tasks in a breadth-first search manner~\cite{javalin}.
As such, the computation pattern of breadth-first search is critical to much of HPC. 
However, the execution of breadth-first search is highly dependent on the graph input, as seen in Section~\ref{sec:results}.
As such, selecting the right self-scheduling method would be difficult, and built-in scheduler methods have been constructed for several HPC breadth-first search implementations that are application, hardware, and input specific~\cite{bcnuma, javalin}. \\

\begin{figure}[tbh]
\centering
\subfloat[\texttt{arabic-2005} in natural ordering]{\label{fig:nat}\includegraphics[scale=.27]{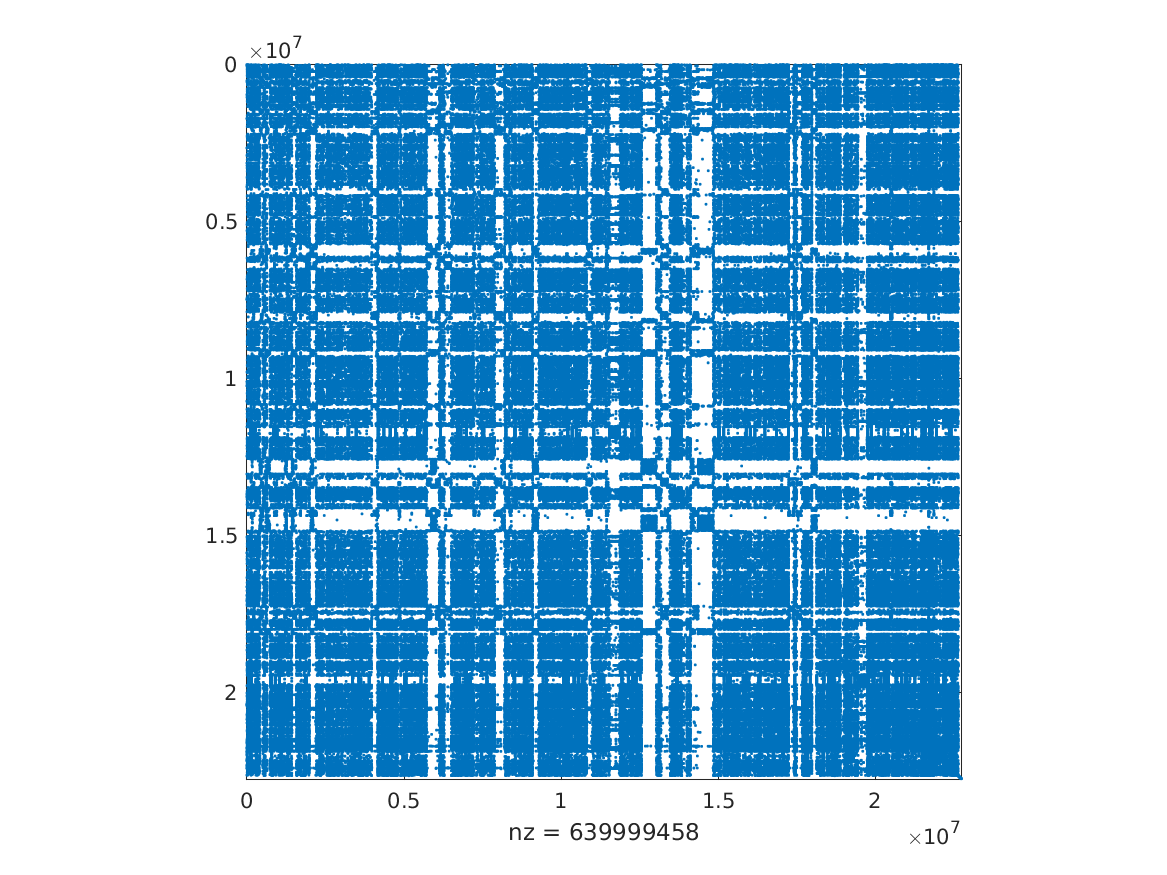}}
\subfloat[\texttt{arabic-2005} in RCM ordering]{\label{fig:rcm}\includegraphics[scale=0.27]{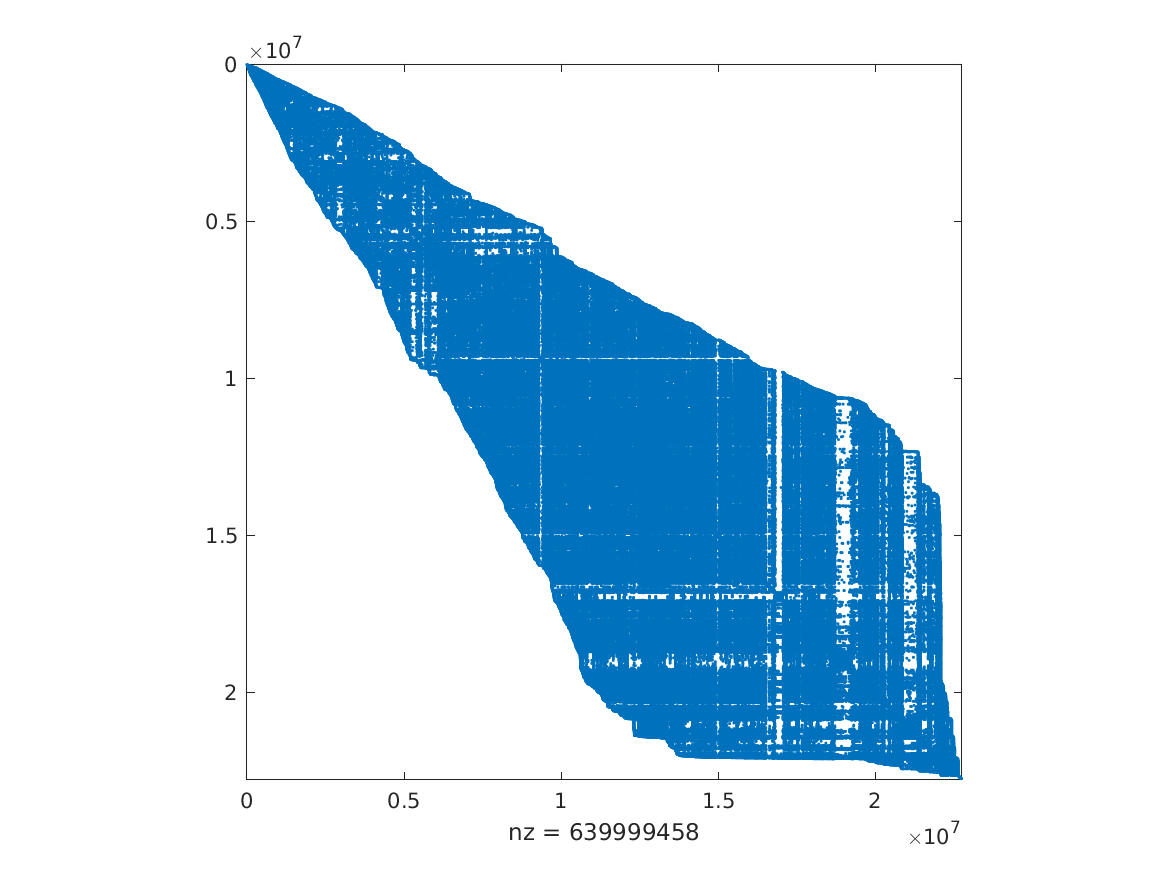}}
\subfloat[Number of rows binned together based on nonzero count in
increments of 50 for \texttt{arabic-2005} (y-axis in log-scale)]{\label{fig:bin}\includegraphics[width=.42\textwidth]{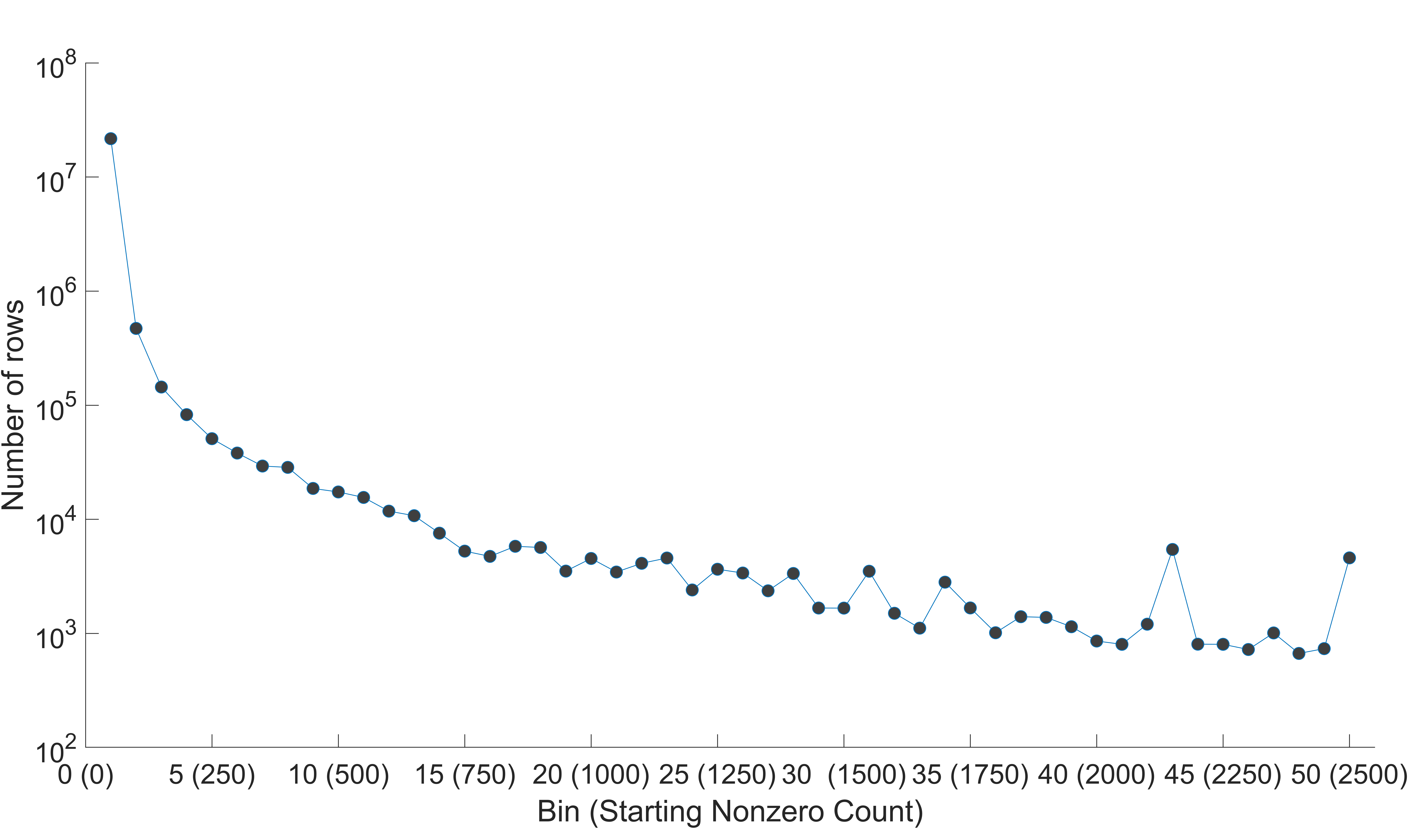}}
\caption{Representations of irregular inputs. Figure (a) presents the sparsity pattern of \texttt{arabic-2005} in its natural ordering. The blue represents nonzero elements and white represents zero elements in the matrix. The natural ordering is the order the comes with the matrix. Figure (b) presents the sparsity pattern of \texttt{arabic-2005} in its RCM ordering.  The RCM ordering is commonly used in sparse linear algebra to permute nonzero elements towards the diagonal in order to improve cache performance. Figure (c) provides the distribution of nonzero elements in the rows of the \texttt{arabic-2005} matrix. In order to make the chart more readable, the distribution has been binned into increments of 50 (e.g., the first dot represents the number of rows with nonzero counts being between 0 and 49).}
\end{figure}

\noindent \textbf{Sparse Matrix-Vector Multiplication.}
The sparse matrix-vector multiplication (\spmv) application is a highly studied and optimized kernel due to its importance in many applications~\cite{Kabir2014,vuducfast, Toledo1997, Pinar1999}.
However, the irregular structure of the sparse coefficient matrix makes this difficult.
If a one-dimensional layout is applied, the smallest task of work is the multiplications of all nonzeros in a matrix row by the corresponding vector entries that are summed together at the end.
Figure~\ref{fig:nat} presents the nonzero structure (i.e., blue representing nonzero entries and white representing zero entries) of the input matrix \texttt{arabic-2005}, which represents the web crawl of sites written in Arabic, in its natural order.
The natural order is the one provided by the input file, and many times this ordering has some relationship to how elements are normally processed or the layout of the system.
From afar, a static assignment of rows may seem like a logical choice.
To investigate, we bin rows based on nonzero counts with increments of 50 together, such that the first bin counts the rows with 0-49 nonzero elements and the second bin counts the total number of rows with 50-99 nonzero elements.
In Figure~\ref{fig:bin}, we provide the tally of the number of rows in each bin (provided in logarithmic scale) for the first 50 bins. 
For example, the first dot shows that there exists close to 21,000,000 rows (e.g., $\sim 95\%$ of rows in the matrix) with nonzero counts between 0 and 49.
We note how much variation exists ($\sigma^2 = 3.0\times10^5$) and the imbalance of work, such as the last two dots representing bin 49 and 50. 
Additionally, matrices are often preordered based on application to provide some structure, such as permuting nonzero elements towards the diagonal or into a block structure.
One such common permutation is the reverse Cuthill-McKee (RCM)~\cite{rcm}.
This little structure can provide some benefits to hand-tuned codes~\cite{Kabir2014,Toledo1997,basker} that can use the newly found structure to better load balance work.
However, Figure~\ref{fig:rcm} shows that this could make balancing even harder if rows were assigned linearly.
Though orderings like RCM may improve execution time~\cite{bcnuma,Kabir2014,rcm}, these orderings may make tuning for chunk size more important. \\

\noindent\textbf{Irregular Application Insight.}
Despite the irregular nature based on input, there does exist some local structure normally.
For example, rows or subblocks within a matrix that have a large number of nonzero elements could be grouped based on some ordering.
The same can also be applied to graph algorithms like breadth-first search. 
Even if the given input does not come with this structure, the input can be permuted to have it.
This type of permutation or reordering is commonly done to improve performance~\cite{bcnuma, Kabir2014, Toledo1997, Pinar1999}. 
Therefore, a thread could, in fact, adapt its own \chunk to fit the local task length.
Moreover, if a thread is finished with its own work, it could be intelligent in stealing work based on the workload of others.
This does require some computation overheads such as keeping track of workload and communicating this workload to nearby neighbors.
Since most of these applications are memory-bound, there does exist a certain amount of availability of computational resources and time during computing.
In particular, Aupy, Benoit, Dai, et al.~\cite{aupy} show that through intelligent co-scheduling of applications performance can be seen not to degrade due to cache overuse, and total performance (i.e., the time to execute all applications) can be improved.
This observation allows us to do some small amount of intelligent record keeping and make smart decisions, and also makes auto-tuning for chunk size even more important (e.g., co-scheduling multiple applications).

\section{Adaptive Runtime Chunk for Irregular Applications}
\label{sec:adapt}
\begin{figure}[tbh]
    \centering
    \includegraphics[width=.99\textwidth]{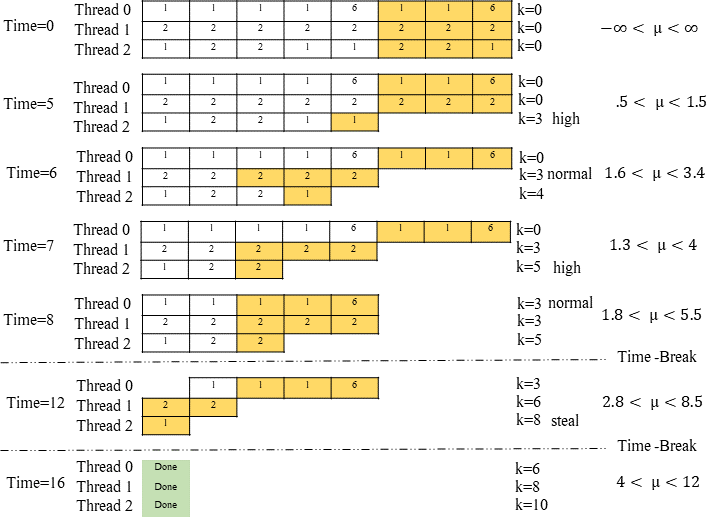}
    \caption{Time step analysis of \ichunk for 3 threads. 
    Each time label is a time step in the scheduler where new chunks of iterations are issued.
    To the right of each time, each thread contains a local queue that is represented by blocks to the right of the thread number. Each block represents a loop iteration, and each number in the block represents the amount of work in that iteration. The blocks shaded yellow are actively being worked on by that thread's chunk. We note that the starting workload balance is: Thread 0 = 18, Thread 1 = 16, and Thread 2 = 12.  The variable $k$ keeps track of the number of iterations successfully executed. A thread that finishes at a given time step will make a decision about its own chunk size relative to a range around the mean ($\mu$).
    This range (i.e., $\mu - \delta \leq \mu \leq \mu + \delta$ where $\delta$ is an estimate of standard deviation) provides insight into the common variance in workload seen by all threads, and a thread can judge if its own iteration count is high, low, or normal compared to this range.
    A decision to steal will result if the local queue is out of work.
    A thread that makes this decision during a time step provides its decision to the right of its own $k$ value in the figure.
    The running range that is used for the decision is provided to the right of each time step.
    Time-Breaks are provided between Time=8 and Time=12 indicating that other time steps exist where threads will finish and make decisions, but these time steps are not included in the figure.
   }
    \label{fig:timesetp}
\end{figure}

This section provides details related to the algorithms and data structures used by \ichunk.
The \ichunk method uses work-stealing as its primary load-balancing mechanic.
Despite the theoretical ability of work-stealing to achieve good load balance, most work-stealing methods require highly optimized queues, and many implementations utilize multiple queues mapped to either processing cores or non-uniform memory access (NUMA) regions~\cite{omptask1,omptask2,kappi}.
These distributed queues help to reduce memory contention when multiple threads try to access a queue simultaneously. 
However, the same contention can happen with distributed queues when the initial partitioning of the workload is uneven. 
The \ichunk method makes this observation about contention at distribution iteration queues and tries to alleviate issues with an adaptive chunk size for each queue.
Therefore, \emph{our new method extends the traditional work-stealing method that uses fixed chunk size by adding an adaptive algorithm for chunk size}.
The adaptive algorithm for chunk size allows for a large enough chunk size to limit the number of times a thread has to access its local queue to dispatch the next active chunk of work but makes the chunk size small enough so that other threads can steal from its local queue without failure because of iterations already being assigned to an active chunk.
As such, \ichunk tries to adjust the chunk size based on a running estimate of the distribution of iteration throughput (i.e., informally, the width that observations span away from the mean).
The following subsections detail the above based on three phases of execution: initialization, local adaption, and remote work-stealing.
Figure~\ref{fig:timesetp} provides a visual overview \ichunk used in the following subsections.

\subsection{Initialization}
Standard methods like \dynamic scheduling in \texttt{libgomp} use a centralized queue and a single chunk size for all threads, but do not scale well with the number of tasks and threads needed to service modern many-core systems.
Therefore, \ichunk uses local queues constructed for each thread that we denote as $q_i$ where $i \in \{0,1, \ldots, p-1\}$ is the thread ID for $p$ threads.
The local structure is memory aligned and is allocated using a first-touch allocation policy containing a pointer to the local queue, local counter ($k_i$), and a variable used to calculate chunk size ($d_i$). 
The tasks are evenly distributed to tasking queues such that $|q_i| = n/p$ where $n$ is the total number of tasks.
Additionally, $k_i = 0$ and $d_i = p$ such that the initial chunk size is $n/p^2$, i.e., $chunk\ size = |q_i|/d_i$. 
The rationale for this choice is that the scheduler wants to allow a chunk size small enough that $p-1$ threads could steal from the queue later.
Moreover, the chunk size becomes smaller as $p$ increases and allows for the variation of tasks that come with more threads.
Figure~\ref{fig:timesetp} shows the initial setup of \ichunk at Time=0. 
We note the iterations, i.e., the blocks, are evenly distributed to the local queues, i.e., the list of blocks to the right of the thread name.
Moreover, the initial chunk size is set to $3 \approx n/p^2$ as shown by the blocks colored yellow. 

\subsection{Local Adaption}
In traditional work-stealing methods\cite{cilk,kappi}, the chunk size is fixed, and any load imbalance is mitigated through work-stealing once all the tasks in the initial queue have been executed.
However, a thread can only steal work that is not already being actively processed, i.e., not in the active chunk of the victim.
Therefore, making the chunk size too large to start will result in a load imbalance that the self-scheduling method may not be able to recover from using work-stealing.
Additionally, making the chunk size too small would result in added overhead and possibly more time to converge.  

The \ichunk method improves upon traditional work-stealing by locally adapting chunk size to better fit the running distribution of iterations completed across threads.
In particular, a thread is classified in regards to its computational load relative to other threads as
\begin{align}
\mbox{low:} \; \; & k_i < \sum_{j=0}^{p-1} k_j/p - \delta,   \\
\mbox{normal:} \; & \sum_{j=0}^{p-1} k_j/p -\delta \leq k_i \leq \sum_{j=0}^{p-1} k_j/p + \delta, \\
\mbox{high:} \; \; & k_i >  \sum_{j=0}^{p-1} k_j/p + \delta . 
\end{align}
This classification into the distribution is based off the standard notation of intervals around the mean ($\mu$), i.e.,
\begin{equation}
    \mu - \delta \leq \mu \leq \mu + \delta
\end{equation}
where $\delta$ is some multiple of the standard deviation ($\sigma$).
In our case, $\mu = \sum_{j=0}^{p-1} k_j/p$, which is the mean number of iterations complete per thread or otherwise called mean iteration throughput. 

However, the question of selecting $\delta$ is of great importance as this will try to capture the standard deviation of the iteration throughput that exists in the irregular application.
For instance, $\mu \pm 2 \sigma$  would provide an interval of two standard deviations around the mean, and $95\%$ of the iterations per thread should be in this interval if they follow a normal distribution. 
This variance or standard deviation is important as iterations may vary greatly in the number of float-point operations and memory requests.
Additionally, a single computational core that is mapped to a local thread can vary in voltage, frequency, and memory bandwidth due to load on the system~\cite{phase}.
As such, even if the workload was uniform among threads, the difference in execution rates between threads may be significantly different.  
Therefore, irregular applications rarely have a normal distribution of iterations per thread. 
Additionally, though $\delta$ is normally a multiple of standard deviation, i.e., 
 \begin{equation}
     \sigma = \sqrt{\frac{\sum (k_i - \mu)^{2}}{p}},
 \end{equation}
the standard calculation would require previous knowledge of the workload.
Though workload-aware scheduling makes use of previous knowledge, information related to voltage and frequency could change from run to run.
A running approximation exists~\cite{runningmean}:
 \begin{eqnarray} 
    \mu_{i+1} &=& \mu_{i} + (k_i - \mu_i)/p \\ \label{eq:var}
    \sigma^{2}_{i+1} &=& \sigma^{2}_{i} + (k_i - \mu_{i})(k_{i}-\mu_{i+1}).
 \end{eqnarray}
 where $i$ represents the time step.
 However, keeping past values of standard deviations and means would be expensive for a lightweight loop scheduler.
 As such, \ichunk estimates this value with a fractional multiplier ($\epsilon$) of the running mean, i.e., 
 \begin{equation}
 \label{eq:delta}
 \delta = \epsilon \sum_{j=0}^{p-1} k_j/p.
\end{equation}
The idea is that variation or standard deviation would be a multiple of the running mean, and this multiplier would provide a way to either tighten (i.e., make the interval smaller) or loosen (i.e., make the interval larger) for the user.
Additionally, $\delta$ will grow with the number of iterations completed. 
As a result of this relationship between $\delta$ and the number of completed iterations, \ichunk is more likely to only adapt chunk size in the beginning due to extremely large variance in the number of completed iterations, and the possibility of adapting due to smaller variance increases with execution. 
This makes sense as \ichunk would have a better global understanding of the workload the longer the application runs, and it would allow more flexibility of chunk size as the local queue become smaller and work-stealing more important.

In \ichunk, the only parameter that the user must select is this $\epsilon$ in equation~\ref{eq:delta}
Additionally, we demonstrate in Section~\ref{sec:results} that \ichunk is not that sensitive to $\epsilon$, allowing for the user to easily pick an $\epsilon$ and still have good performance.
This observation allows \ichunk to be used across different applications, systems, and inputs without hand-tuning by the user to achieve ``good" speedups.

Once the thread is classified into its category of low, normal, or high, the chunk size of the thread can be adjusted.
As noted previously, $d_i$ is used to directly adjust the chunk size, i.e., $\chunk = |q_i| / d_i$.
After classification, $d_i$ is adjusted as follows.
If the thread is classified as low, then $d_i = d_i/2$, and the chunk size would increase (i.e., $\chunk   = |q_i| / (d_i/2) = 2 |q_i| / d_i$).
If the thread is classified as high, the chunk size would decrease by allowing $d_i = 2 \times d_i$.
Lastly, if the thread is classed as normal, $d_i$ would remain the same.
The logic for the direction of updating the values of $d_i$ is as follows.
If a thread is classified as low, then it is completing fewer iterations than the mean across all threads.
This could be for several reasons, but one likely reason is that the chunk size is already too small and the thread is spending too much time assigning the next chunk.
One other reason is that the thread is processing iterations slower, e.g., lower frequency or processing memory requests.
We note that even though the workload is normally thought of in metrics such as the number of float-point operations, the time of irregular applications is dominated by memory operations, communication, and scheduling overheads, and the time is not dominated by the irregular proportion of float-point operations. 
Therefore, \ichunk will assign a larger chunk size so that this thread will be less likely to have to deal with interruptions due to scheduling or stealing.
On the other hand, a thread classified as high is completing iterations faster than the thread.
This thread has time to spare in updating chunk size, scheduling, or dealing with stealing operations, and therefore, \ichunk makes the thread chunk smaller.
We note that the direction in which chunk size is updated is in opposition to the logic that one might have if optimizing to balance the average amount of work assigned to each thread in a chunk, where a thread classified as high would have its chunk size increased and a thread classified as low would have its chunk size decreased to attempt to approach the mean time spent processing a chunk across all cores.

In Figure~\ref{fig:timesetp}, the chunk size adaptation can be observed. 
At Time=5, only Thread 2 is done with its work, thus is running faster than the other threads.  
Thread 2 reduces its chunk size by half. 
This is possible since it is faster, and therefore, more time can be spent dealing with coming back to the thread queue to leave more iterations that can be stolen by other threads.
By Time=6, Thread 1 observes that it is running within the appropriate amount of time, i.e, the bound around $\mu$ on the far right-hand side, and so it does not change its chunk size.
This trend continues throughout the time steps.

\subsection{Remote Work Stealing}
At some point, the local queues will start to run out of work, and then work-stealing is used, as in Time=12 in Figure~\ref{fig:timesetp}.
Many implementations of work-stealing methods will fix a chunk size and use the THE protocol~\cite{cilk,kappi}.
In Listing~\ref{fig:thee}, we provide the THE protocol's stealing algorithm updated with \ichunk's needed additional calculations. 
The THE protocol tries to steal from a random victim (\texttt{victim}) half of its queue (line 4).
If stealing is unsuccessful (e.g., the victim's beginning (\texttt{victim->begin}) is updated before stealing is finished by the victim (line 12) or some other stealing thread) then it rolls back its attempt while trying to minimize the number of locks required.
Note that the \texttt{victim} is only locked after queue size and other important variables are calculated (line 9), and the \texttt{victim} is unlocked as soon as possible (line 15 or 18).
Additionally, the \texttt{thief} does not have to lock itself at all, and all this is set up to try to minimize contention dealing with locks.
Unlike the traditional method that has a fixed chunk size, \ichunk must deal with the issue of assigning a chunk size to the stealing thread and does not know if the victim's thread and stealing thread's current chunk sizes were due to characterizing the workload or performance of the thread on the particular computational core.
Therefore, the stealing thread's  $d_i$ and $k_i$ are updated based on the victim's $d_j$ and $k_j$ by taking the average of both, i.e., $d_i = (d_i + d_j)/2$ and $k_i = (k_i + k_j)/2$ (line 6 and 7).
The reasoning for this choice is as follows:
the stealing thread knows some information from the victim;
however, the stealing thread does not know the accuracy of the information, so it tries to average out the uncertainty with its own knowledge.

\begin{centering}
\begin{lstlisting}[caption={Remote stealing from victim with the THE protocol. Given the \texttt{victim} and \texttt{thief} structures containing the queue's current \texttt{begin}, \texttt{end}, \texttt{ki} ($k_i$), \texttt{di} ($d_i$), and thread lock.}, label=fig:thee, frame=single, basicstyle=\footnotesize, numbers=left]
//Test if iterations still exist to steal from victim
if(victim->end - victim->begin >0){
    //Calculate half of the remaining iterations from the victim
    halfsize = (victim->end-victim->begin)/2;
    //Calculate the average k_i and d_i of victim and thief
    localiter = (victim->ki + thief->ki)/2;
    localchunk = (victim->di + thief->di)/2;
    //lock victim
    lock(victim)
    end = victim->end - halfsize;
    victim->end = end;
    if(end <= victim->begin){
        //Abort --- Rollback if cannot steal half the remaining
        victim->end = end + chunksize;
        unlock(victim);
        return false;
    }
    unlock(victim);
    //Make sure new chunk size is viable
    if(halfsize <= localchunk){
        localchunk = halfsize;
    }
    thief->begin = end;
    thief->end = end + halfsize;
    thief->ki = localiter;
    thief->di = localchunk;
    return true;
}
return false;
\end{lstlisting}
\end{centering}


\section{Related Work}
\label{sec:relatedwork}
Subramanian and Eager~\cite{equal} introduce an affinity loop scheduler for unbalanced workloads for a series of parallel loops if ``execution time of any particular iteration does not vary widely from one execution of the loop to the next''. 
Their work uses multiple loop executions to learn to balance the workload, and later work builds around this fundamental work.
However, \ichunk tries to forge in the new direction of not using multiple loops and recognizes the variable nature of modern systems. 
Work by Yan, Jin, and Zhang~\cite{yan} adds a dynamic history to decide about load-balance on distributed shared-memory systems.
The adaptive algorithm in \ichunk is inspired by their work.
They both use running sums of iterations and increase chunk size by a factor of 2.
However, they attempt to optimize for load-balance, and most of their classification end up being the opposite of those in \ichunk (i.e., because \ichunk tries to optimize for differences in iteration throughput that might affect stealing). 
Their work considers how to keep and update a history on distributed shared-memory systems, such as KSR-1 and Convex up to 16 CPUs, that could have high delays and a different memory architecture compared to today's modern systems.
They test their method on the applications of Jacobi iteration method, transitive closure, and successive over-relaxation, matrix-matrix multiplication, and adjoint convolution.
However, both Jacobi iteration method and successive over-relaxation reduce to \spmv, and the input sparse matrix they use is a banded matrix with a somewhat regular structure.
Yan does not compare against any modern self-scheduling method but demonstrates low overhead and about a $4\times \sim 5.5 \times$ speedup using their method on 16 CPUs.

Additionally, more modern irregular loop methods are studied~\cite{factoring,binlpt, history, wang2012}.
Banicescu, Velusamy, and Davaprasd~\cite{factoring} schedule loops of irregular applications in a distributed fashion with MPI.
The chunk size is determined by a direct fraction (i.e., a \emph{factoring self-scheduling} method) of the cumulative number of tasks complete and the processor speed, called Adaptive Weighted Factoring (AWF).
However, AWF suffers from needing to know the processor speed, and chunk size is updated based on its history rather than by a fixed amount in \ichunk.
To evaluate AWF, they use a Laplace Equation Solver using Jacobi iterations and an N-Body Simulation and show up to a $45\%$ improvement over standard static scheduling.
Wang, Ji, Shi et al.~\cite{wang2012} introduce the KASS system that considers adaptively chunking together in the initialization phase using some prior knowledge (i.e., \emph{workload-aware self-scheduling}). 
Chunks in the second phase (i.e., the dynamic or adaption phase) are reduced fixedly based on information from past iteration runs but are not adapted (i.e., chunk size can not increase, reduce, or stay the same) within an iteration, unlike the chunk size in \ichunk.
Chunks are stolen if a queue runs out of its own work.
To evaluate KASS, they use in-house versions of successive over-relaxation, Jacobi iteration method, and transitive closure. 
Like the work by Yan, both Jacobi iteration method and successive over-relaxation reduce to \spmv, and the input sparse matrix they use is a banded matrix with a somewhat regular structure.
However, KASS is only able to obtain $21\%$ improvement to their baseline of \emph{guided self-scheduling} (i.e., a method similar to OpenMP \guided). 
Kejariwal and Nicolau~\cite{history} introduce a \emph{history-aware self-scheduling} method, called HSS, that changes chunk size from past iterations and the number of times the task will run using a complex and expensive ``best-fit"  approximation fit.
Compared to HSS, the \ichunk method uses a simple and cheap method to adapt the chunk size.
HSS is evaluated on irregular loops extracted from the Standard Performance Evaluation Corporation (SPEC) Benchmarks~\footnote{Available at www.spec.org} and is tested against AWF.
Using a simulator (i.e., not a real system), HSS outperformed the baseline by up to $18\%$.
Their method does demonstrate benefits for loops that are repeated, but \ichunk does not consider this as the loops may not vary within an irregular application.
Lastly, BinLPT~\cite{binlpt} schedules irregular tasks from a loop using an estimate of the work in each loop and a maximal number of chunks provided by the user.
The method uses this information to distribute chunks of iterations to local queues that are theoretically balanced.
If there is unbalance a simple \emph{chunk self-scheduling} method is used to balance and provides little scheduling overhead.
A multiple loop analysis has been added to BinLPT so that it may better distribute loop iterations when a loop is repeatedly called.
This method is one of the newest and provides good performance in publication.
In evaluating BinLPT, a synthetic benchmark~\footnote{Available at www.github.com/lapese/libgomp-benchmarks}, \spmv, minimal spanning tree built around Prim's algorithm, and an N-Body simulation for computational fluid dynamics are used.
BinLPT compares against in-house versions of KASS and HSS.
On relatively few threads (e.g., $< 48$), the performance of BinLPT is about as good as \guided.  
However, BinLPT is able to outperform \guided when using more threads.
As such, \ichunk is tested against BinLPT.
In contrast to BinLPT, the  \ichunk method provides an easier choice that does not require estimates of loop work and more user input.
No other self-scheduler work directly attacks the problem of providing a method that works well on all irregular applications.

\section{Experimental Setup}
\label{sec:exp}
\subsection{Test Applications}
Test applications are selected that reflect the applications used by other self-scheduling methods to evaluate. 
Additionally, other applications are introduced for testing that better define the limits and multiple behaviors that can be found in irregular applications.\\

\begin{figure}[tbh]
\centering

\subfloat[Uneven distribution in a particle simulation.]{\label{fig:spread}\includegraphics[width=.48\textwidth]{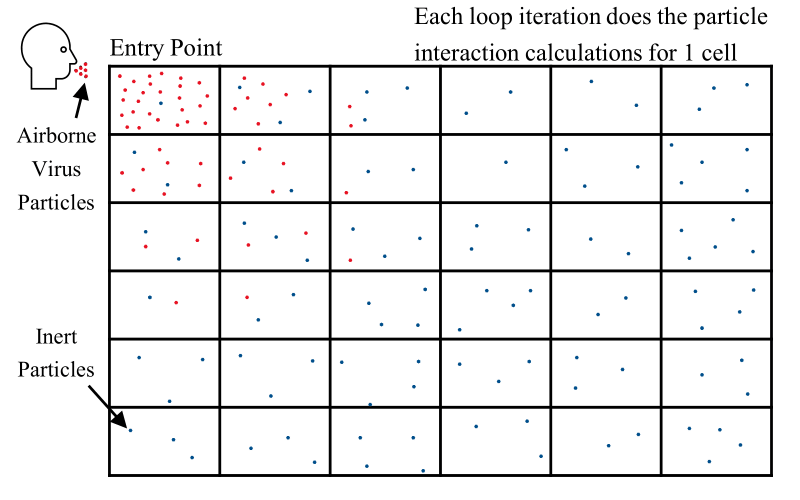}}
\hspace{10pt}
\subfloat[Exponential distribution used for Exp-Decreasing (pictured) and Exp-Increasing (sorted in opposite order as pictured).]{\label{fig:expbin}\includegraphics[width=.45\textwidth]{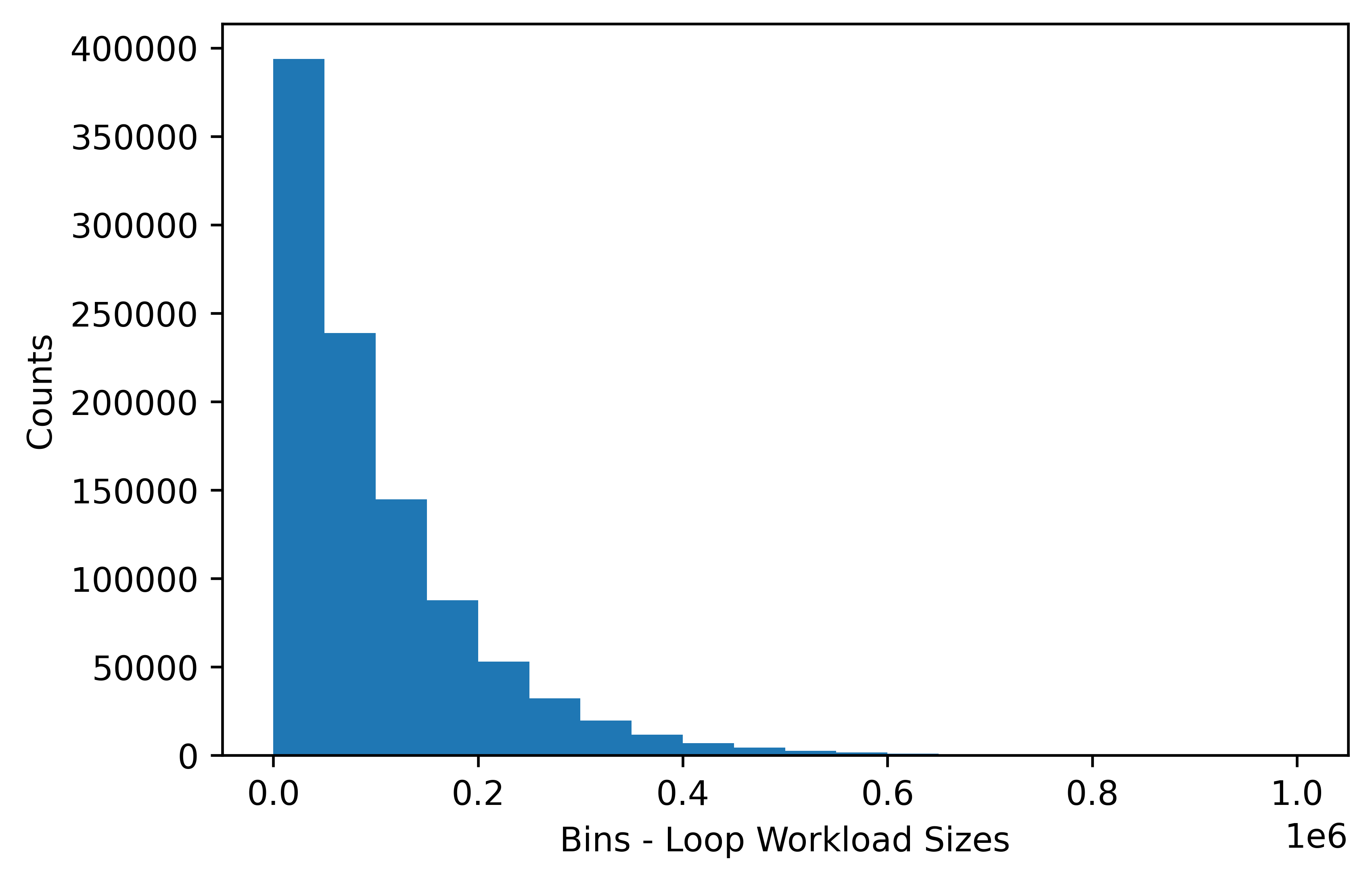}}

\caption{Examples of distributions for \synth and \lava.}
\end{figure}

\noindent \textbf{Synth (\synth)}.
This application is a synthetic benchmark used by BinLPT~\cite{binlpt} to evaluate their performance.
This benchmark allows the user to input a custom workload distribution.
In BinLPT~\cite{binlpt}, the benchmark is ran with linear, logarithmic, quadratic, and cubic workload distributions, but none of their testings demonstrates any significant performance variation between these distributions.
Therefore, we use two distributions built around the exponential distribution to demonstrate exponential increase and decay in workloads.
This distribution is more representative of workloads that are highly imbalanced when the loop either starts or ends.
An example of this type of distribution would be tracking particles in an aerosol as they enter a simulation domain that is evenly distributed, such as coughing in a room.
The iterations that compute simulation work near the entry location will complete more computations than iterations tracing particles on the far side of the simulation domain, e.g., the particles in the two-dimensional simulation in Figure~\ref{fig:spread}. 
In particular, 1,000,000 samples from the the exponential distributions (i.e., $pdf(x, \frac{1}{\beta}) = \frac{1}{\beta} exp(\frac{-x}{\beta}))$ are generated with $\beta = 1,000,000$ and sorted.
Entries sorted in increasing order are labeled Exp-Increasing, and those sorted in descending order are labeled Exp-Decreasing. 
We note that the range of loop workload is therefore 1,000,000 to 1.
Figure~\ref{fig:expbin} provides a histogram of the values in the exponential distribution. \\

\noindent \textbf{Breadth-first search (\bfs)}. 
The \bfs in the Rodinia~\cite{rod1} benchmark suite for heterogeneous computing is used.
We use two different inputs.  
The first input is generated using the graph generator provided that uses the uniform distribution to randomly generate the number of neighborhood vertices (Uniform).
The second input is generated using a modified graph generator that uses a power-law distribution to randomly generate the number of neighborhood vertices (Scale-Free) in order to construct a randomly generated scale-free network.
Scale-free networks are those with the fraction of $P(k)$ nodes in the network having $k$ connections is $P(k) \sim k^{\gamma}$ where $\gamma = 2.3$ in our tests.
Scale-free networks are of particular importance as many common networks, such as social networks, computer networks, protein-protein interactions, etc., are scale-free. 
Additionally, \bfs tends to be the building block kernel in many scalable implementations of graph analytic~\cite{bcnuma} commonly used on these networks and is the building block of other graph benchmarks used by others, such as minimum spanning tree with Prim's algorithm tested in BinLPT.\\

\noindent \textbf{K-Means (\kmeans)}.
A \kmeans benchmark, as used in machine learning applications, is used from the Rodinia Benchmark.
In particular, this version uses the KDD Cups dataset related to network packets, and this dataset is the standard in most \kmeans benchmarks.
As a popular method in machine learning, this application is used by many in HPC. 
This benchmark is highly irregular, and the iteration workload distribution in the innermost loop changes per outermost loop iteration. 
This particular benchmark should thus be difficult for \emph{history-aware} methods.\\

\noindent \textbf{LavaMD (\lava)}.
This application is a Computational Fluid Dynamics (CFD) code that performs N-Body Simulations from the Rodinia Benchmark.
LavaMD simulates the interactions of solidification of molten tantalum and quenched uranium atoms in a finite three-dimensional domain. 
We use an input size of $8 \times 8 \times 8$ to construct the domain.
At each step, force calculations are done for particles within the same box. 
However, the simulation is set up to have a cut-off radios of particles interaction at about the size of a box.  
As such, little to no calculations are done between boxes. 
Additionally, particles move relatively small distances, unlike our cough example above in Figure~\ref{fig:spread}.
This benchmark is also used by BinLPT in their evaluation.\\

\noindent \textbf{Sparse Matrix-Vector Multiplication (\spmv)}.
This algorithm is commonly used by sparse linear solver methods (e.g., successive over-relaxation and Jacobi iterative method), and a great deal of work goes into optimizing it for specific hardware and inputs~\cite{Kabir2014}.
Most other self-scheduling evaluations use \spmv in some form within a solver.
The choice of using it within a solver allows there to be a nested loop to add history.
Additionally, other evaluations tend to focus on a single input matrix with a regular structure.
For our inputs, we use a collection of sparse matrices from the SuiteSparse Collection~\cite{matrix}.
Table~\ref{tab:mat} contains the sparse matrices, where the number of vertices and edges are reported in millions (i.e., $10^6$).
Inputs are picked due to their size, variation of density, and application areas.
In particular, four application areas are of particular interest.
These areas are: 
Freescale: a collection from circuit simulation of semiconductors; 
DIMACS: a collection from the DIMAC challenge that is designed to further the development of large graph algorithms;
LAW: a collection of laboratory for web algorithms of web crawls to research data compression techniques;
and GenBank: a collection of protein k-mer graphs.
Furthermore, we report the average row density ($\bar{x}$), the ratio of the maximal number of outgoing edges for a vertex over the minimal number of outgoing edges for a vertex ($ratio$), and the variance of the number of outgoing edges ($\sigma^2$) for each input.
These numbers provide a sense of how sparse and how uneven the work is distributed per vertex.
Some inputs are very balanced, such as input I8 (\texttt{hugebubbles}).
Others have more variance like input I12 (\texttt{uk-2005}). \\

\begin{table}[tbh]
\centering
\caption{Input Graphs. Vertex and edge counts in millions. $\bar{x}$: average number of outgoing edges per vertex. $ratio$: maximal number of outgoing edges over minimal number of outgoing edges. $\sigma^2$: variance of the number of outgoing edges.}
\label{tab:mat}
{
\begin{tabular}{ | l | l | l | l | l | l | l | }
\hline
	\textbf{Input} & \textbf{Area} & $|V|$ & $|E|$ & $\bar{x}$ & $ratio$ & $\sigma^2$   \\ \hline
	I1: FullChip & Freescale & 2.9 & 26.6 & 8.9 & 1.1e6 & 3.2e6 \\ \hline
	I2: circuit5M\_dc & Freescale & 3.5 & 14.8 & 4.2 & 12 & 1 \\ \hline
	I3: wikipedia & Gleich & 3.5 & 45 & 12.6 & 1.8e5 & 6.2e4 \\ \hline
	I4: patents & Pajek & 3.7 & 14.9 & 3.9 & 762 & 31.5 \\ \hline
	I5: AS365 & DIMACS & 3.7 & 22.7 & 5.9 & 4.6 & 0.7 \\ \hline
	I6: delaunay\_n23 & DIMACS & 8.3 & 50.3 & 5.9 & 7 & 1.7 \\ \hline
	I7: wb-edu & Gleich & 9.8 & 57.1 & 5.8 & 2.5e4 & 2.0e3 \\ \hline
	I8: hugebubbles-10 & DIMACS & 19.4 & 58.3 & 2.9 & 1 & 0 \\ \hline
	I9: arabic-2005 & LAW & 22.7 & 639.9 & 28.1 & 5.7e5 & 3.0e5 \\ \hline
	I10: road\_usa & DIMACS & 23.9 & 57.7 & 2.4 & 4.5 & 0.8 \\ \hline
	I11: nlpkkt240 & Schenk & 27.9 & 760.6 & 27.1 & 4.6 & 4.8 \\ \hline
	I12: uk-2005 & LAW & 39.4 & 936.3 & 23.7 & 1.7e6 & 2.7e6 \\ \hline
	I13: kmer\_P1a & GenBank & 139.3 & 297.8 & 2.1 & 20 & 0.4 \\ \hline
	I14: kmer\_A2a & GenBank & 170.7 & 360.5 & 2.1 & 20 & 0.3 \\ \hline
	I15: kmer\_V1r & GenBank & 214 & 465.4 & 2.1 & 4 & 0.3 \\ \hline
\end{tabular}
}
\end{table}

\subsection{Self-Scheduling Methods}
Table~\ref{tab:scheduling} provides a summary of the scheduling methods used in the result section.
Two commonly used self-scheduling methods in OpenMP, i.e., \guided and \dynamic, are tested.
Additionally, the newly added \taskloop~\footnote{https://developers.redhat.com/blog/2016/03/22/what-is-new-in-openmp-4-5-3/} scheduling method is tested as it has provided good results in several cases, and is an active research area for irregular applications.
The workload-aware method from BinLPT~\cite{binlpt} (\binlpt) is also compared against.
We do not compare against HSS and KASS as no source is available, but an in-house version of it was tested against \binlpt, and HSS and KASS were shown to perform worst than \binlpt.
The chunk sizes for \dynamic, \guided, and \binlpt are the same used by BinLPT.
Additionally, the work-stealing method (\stealing) that \ichunk is based on is evaluated to demonstrate the improvement that an adaptive chunk size makes to the base algorithm.
Lastly, the \ichunk method is evaluated with three levels of $\epsilon$ (i.e., $25\%, 33\%, 50\%$).

    \begin{table}[tbh]
        \centering
         \caption{Scheduling Methods used to test.  Parameters are based on those used by the BinLPT~\cite{binlpt}}
        \label{tab:scheduling}
        \begin{tabular}{l|l} \hline
            \textbf{Scheduling Method} &  \textbf{Parameters} \\ \hline
            \guided                    &   chunk size $=\{1,2,3\}$ \\
            \dynamic                  &   chunk size $=\{1,2,3\}$ \\
            \taskloop                  &   num\_task = num\_threads \\
            \binlpt                    &   chunk size $=\{128,384,576\}$ \\
            \stealing                 &   chunk size $=\{1,2,3,64\}$ \\
             \ichunk                   &   $\epsilon = 25\%, 33\%, 50\%$ \\
              \hline
        \end{tabular}
       
    \end{table}

\subsection{Test system}
Bridges-RM at Pittsburgh SuperComputing Center~\cite{xsede} is used for testing. 
The system contains two Intel Xeon E5-2695 v3 (Haswell) processors each with 14 cores and 128GB DDR4-2133.
Other microarchitectures, such as Intel Skylake, are also tested, but results do not vary much.
We implement \ichunk inside of GNU \emph{libgomp} under the GPL v3 License.
All codes are compiled with GCC 8.2.0 except \taskloop.
All \taskloop applications are compiled with GCC 9.2.0 as the \synth application requires a reduction operator that is not implemented in GCC 8.2.0. 
OpenMP threads are bound to cores with \texttt{OMP\_PROC\_BIND=true} and \texttt{OMP\_PLACES=cores}.

\section{Results}
\label{sec:results}
In this section, we break the results into two subsections.
The first subsection focuses on the best general speedup that is obtainable by our test methods.
The goal of the first subsection is to provide a high-level view of how well each of the self-scheduling methods could speed up the wide range of applications in this paper if some minor tuning of parameters, such as chunk size, is done.
In the first subsection, we report our metrics using the best time over the tuning parameters, such as chunk size, listed in Table~\ref{tab:scheduling}.
The second subsection will focus on parameter selection of \ichunk. 
In particular, it will show the sensitivity of \ichunk to parameter selection and compare the impact of parameter selection. 

\subsection{Speedup}
We define $T(app,schedule,p)$ as the best time of running application ($app$) using the self-scheduling method ($schedule$) on $p$ threads across all method parameters in Table 2. 
Additionally, we define as
\begin{equation}
speedup(app,schedule,p) =  \frac{T(app,guided,1)}{T(app,schedule,p)}.
\end{equation}

\begin{figure}[tbh]
\centering
\includegraphics[width=0.95\textwidth]{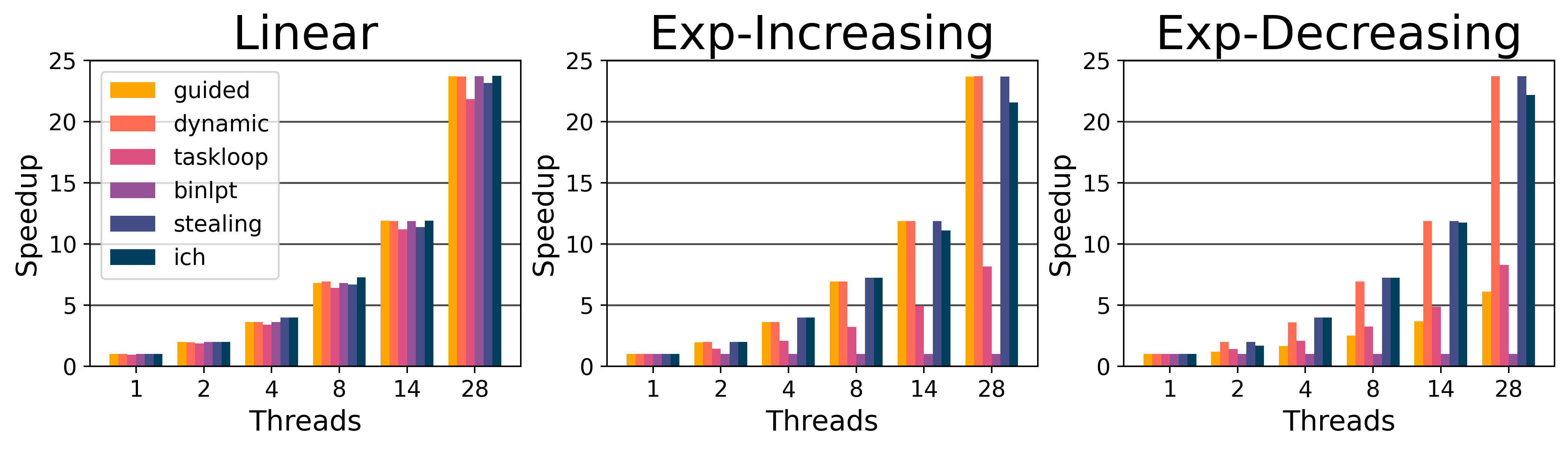}
\caption{Synth Speedup. The speedup for a linear distribution (Linear), exponential increasing distribution (Exp-Increasing), and exponential decreasing distribution (Exp-Decreasing). We observe that \ichunk performs at about the same as the best method for all three inputs. However, we notice that no method (besides \dynamic, \stealing, and \ichunk) does well across all three inputs.}
\label{fig:synth}
\end{figure}

\noindent \textbf{Synth.}
Figure~\ref{fig:synth} presents the speedup of the \synth application.
The first subplot presents the linear workload, which is used in BinLPT~\cite{binlpt}.
In this workload, we see that all self-scheduling methods perform about equally as well, and this finding matches that by BinLPT.
Using 28 threads, we notice that \taskloop is the worst-performing.
The next subplot presents the exponential increasing order (Exp-Increasing).
This means that the iteration workload increases, and therefore, the first threads that are scheduled will have the lightest workloads.
In this case, \guided, \dynamic, and \stealing are the best self-scheduling methods, and they are about the same in terms of speedup.
However, we notice that \binlpt no longer has good performance and \taskloop struggles. 
The \ichunk method is able to achieve a speedup close to the \guided, \dynamic, and \stealing.
Since \ichunk is built around \stealing, and \stealing is one of the best self-scheduling methods for this application and input workload, this difference in the speedup of \stealing and \ichunk provides insight into how much overhead the chunk size adjustment can have, which is $\sim 8.9\%$ in terms of speedup.
In the last subplot (Exp-Decreasing), the iteration workload decreases, and therefore, the first threads that are scheduled will have the heaviest workloads.
The self-scheduling methods of \dynamic and \stealing still perform well.
However, \guided no longer does well and is even worse than \taskloop.
The performance of \guided is due to how \guided assigns the largest chunk sizes of work to the first threads, and the small chunk sizes later will not be able to rebalance the workload among the threads. 
The \ichunk method again is not the best but is close (i.e., $\sim 6.5\%$ difference in speedup from \stealing) to the best, indicating that the overhead of adapting the chunk size has an overhead, but does not dramatically change performance.\\

\begin{figure}[tbh]
\centering
\subfloat[Breadth-first Search Speedup]{\label{fig:bf}\includegraphics[width=.628\textwidth]{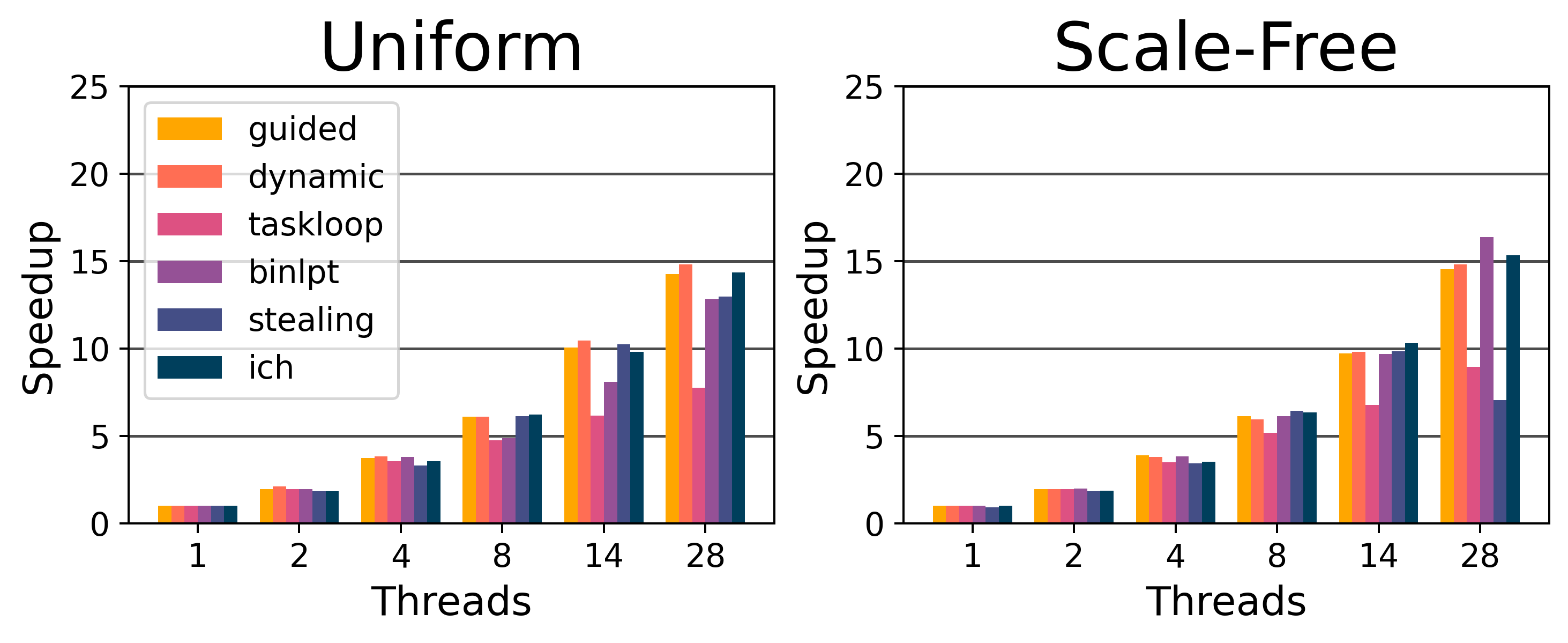}}
\subfloat[K-Means Speedup]{\label{fig:kmeans}\includegraphics[width=.32\textwidth]{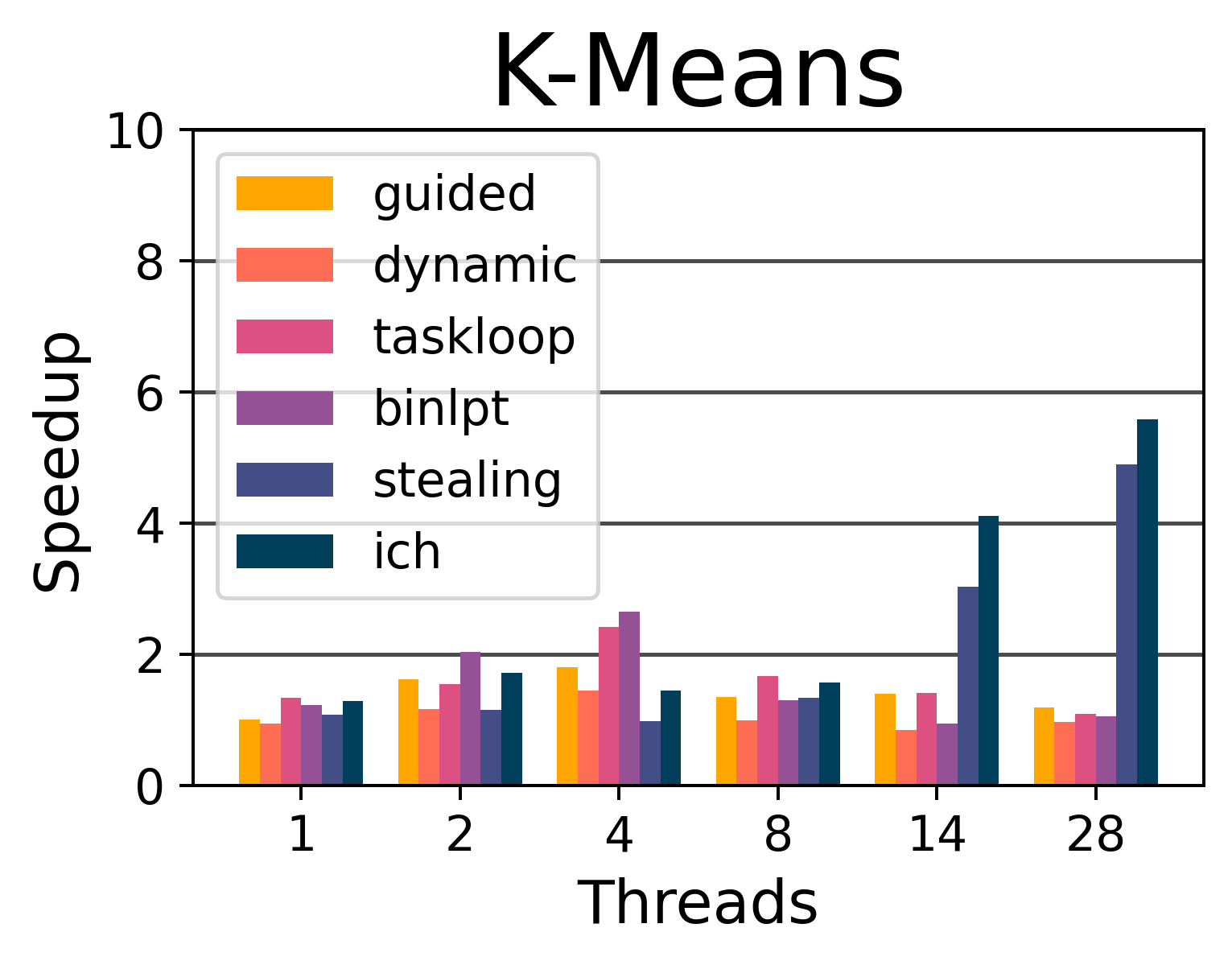}}\\

\caption{Speedup for Breadth-first Search (a) and K-Means (b).  Breath-first search is tested with two inputs: Uniform and Scale-Free. We observe in (a) that \ichunk does better than the \stealing method alone that is used by \ichunk to help with imbalance. In (b), we observe that only \stealing and \ichunk can scale past 8 threads, and that \ichunk is again able to outperform \stealing alone.}
\end{figure}

\noindent \textbf{Breadth-first search.}
Figure~\ref{fig:bf} presents the speedup of the \bfs application.
The first subplot (Uniform) presents the speedup of \bfs applied to the graph when the number of neighboring vertices generated from a uniform distribution.
In this case, the best scheduling methods are \dynamic, \ichunk, and \guided, in that order.
We notice that \stealing does not perform as well as \ichunk even though it is the base algorithm for \ichunk, i.e., \ichunk is $\sim 9.6\%$ better in terms of speedup than \stealing.
This observation helps to demonstrate this work's conjecture that work-stealing based self-scheduling could be improved by having a chunk size that will not interfere with stealing.
The \binlpt method does about as well as \stealing, and \taskloop does the worst.
The second subplot (Scale-Free) presents the speedup when \bfs is applied to a scale-free graph, i.e., the number of neighboring vertices follows the power-law distribution. 
In this case, \binlpt and \ichunk have the best speedups.
We note that the speedup for \dynamic and \guided did not generally go down, but \binlpt and \ichunk were just better able to improve the speedup.
For \binlpt, the performance increase seems to be due to the secondary self-scheduling method, which is based on \guided, performing better.
For \ichunk, the performance was due to less overhead in chunk size adaption (i.e., any overhead helped to yield better performance). 
More importantly, we observed that \ichunk does much better than \stealing (i.e., \ichunk is $\sim 54\%$ better than \stealing), and demonstrates that the addition of adaptively changing chunk size pays off in performance.\\ 

\noindent \textbf{K-Means.}
Figure~\ref{fig:kmeans} presents the speedup of the \kmeans application.
This application is known for not scaling well as the iteration workload is very uneven, and the workload is changed for every outermost iteration.
This makes any \emph{history-based self-scheduling} method that tries to learn the workload from a past iteration not useful, and the only information that can be used is the information from that iteration itself. 
Because of its highly irregular nature, the only self-scheduling methods that can obtain any real speedup are \ichunk and \stealing.
For thread counts up to 8, all self-scheduling methods perform equally as well.  
Based on experimenting with different thread placements, this is due to memory pressure when threads are placed close and share cache space.
However, when a larger number of threads are used, such as 14 and 28, the memory pressure is reduced and \ichunk and \stealing can continue to speed up while the other methods do not continue to scale.
In this application, we again see that \ichunk is able to outperform the standard \stealing method by $\sim 12.3\%$ in terms of speedup. \\

\begin{figure}[tbh]
\centering
\subfloat[LavaMD Speedup]{\label{fig:lava}\includegraphics[width=.45\textwidth]{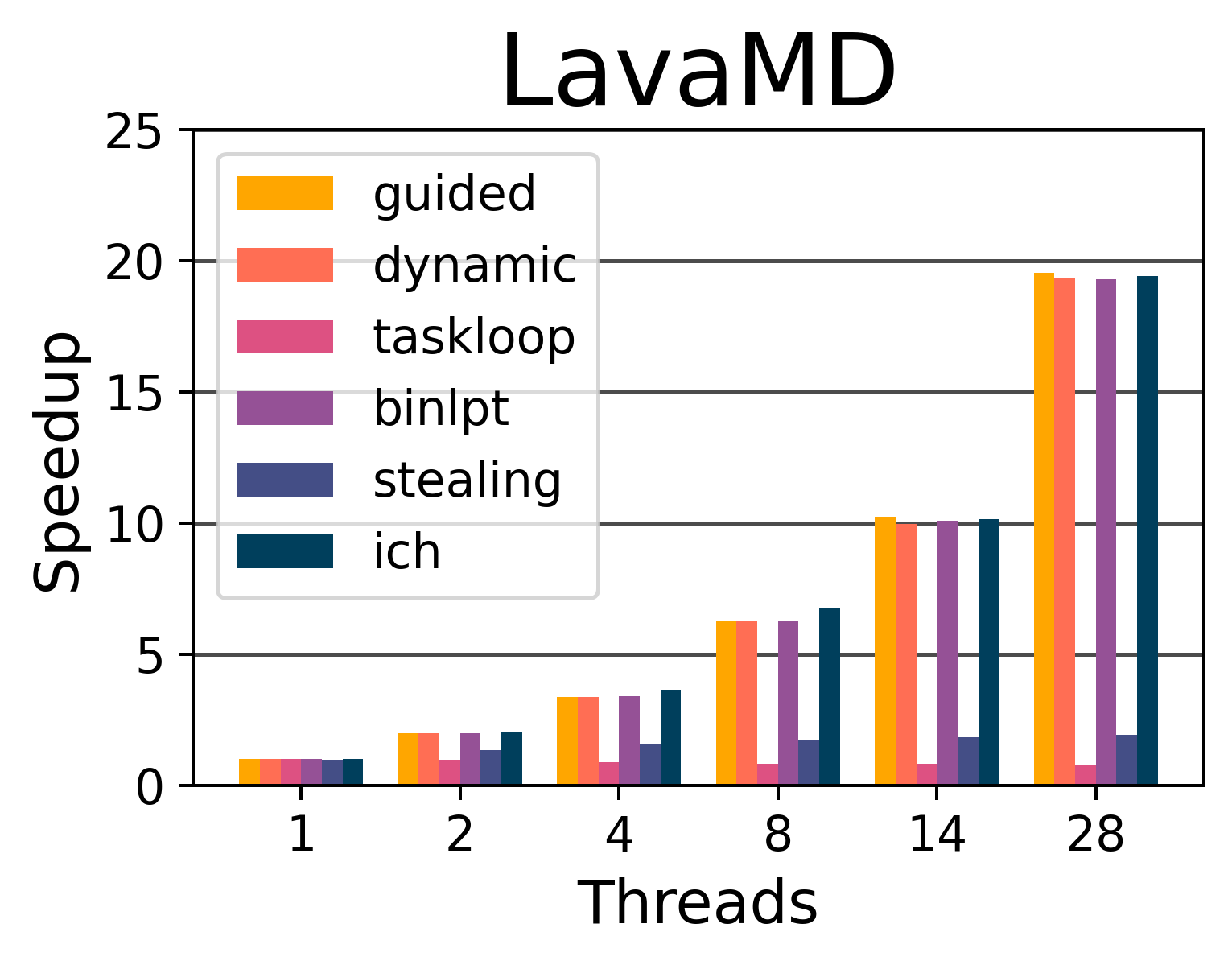}}
\subfloat[Spmv Speedup]{\label{fig:spmv}\includegraphics[width=.45\textwidth]{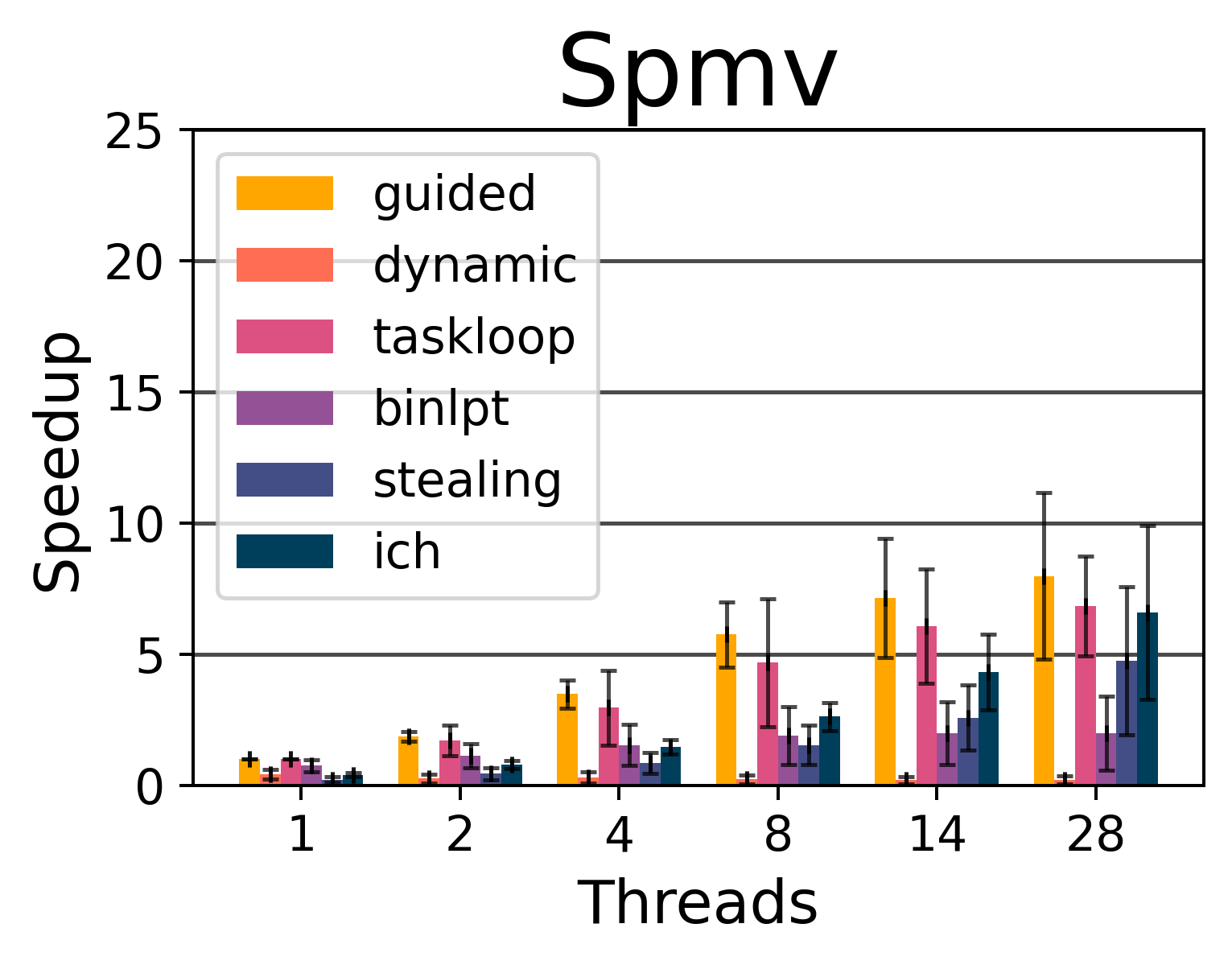}}\\
\caption{Speedup for LavaMD and Spmv. For LavaMD (a), most self-scheduling methods are about equal except \taskloop and \stealing. This demonstrates the improvement that \ichunk's adaptive chunk size has over the base algorithm of \stealing.  }
\end{figure}

\noindent \textbf{LavaMD.}
Figure~\ref{fig:lava} presents the speedup of the \lava application.
Though this application is a molecular-dynamics application, the iteration workload is relatively well balanced. 
We observe that almost all self-scheduling methods do about the same except for \stealing and \taskloop.
Both \stealing and \taskloop are unable to scale at all.
For \stealing, this seems to be related to the few numbers of loops (i.e., $512 = 8 \times 8 \times 8$)  and the relatively few chances \stealing has to make up for: (a) trying to steal from a thread that is locked up due to chunk size and (b) randomly selecting from nonoptimal choices once the stealing thread tries to recover from the issue (a).
However, the \ichunk method can obtain speedup matching that of \guided and \dynamic.
This is due to \ichunk quickly setting a chunk size that would allow threads with small workloads to steal in order to rebalance.
This again demonstrates \ichunk's adaptive chunk size improving the base implementation of \stealing.\\

\noindent \textbf{Sparse Matrix-Vector Multiplication.}
The performance of the \spmv application varies greatly based on input. 
As pointed out in the previous sections, a large variety of spare input matrices are tested in order to obtain the wide range of performance of \spmv, and most \spmv like applications used to evaluate other self-scheduling methods tend to only focus on one sparse matrix with a somewhat regular structure.
Figure~\ref{fig:spmv} presents the geometric mean of speedups as bars.
Additionally, the best and worst speedup for each scheduling method on a particular thread count across all sparse matrix inputs is presented as upper and lower whiskers of the bars. 
Over the whole test set, we observe that \guided has the best speedup with \taskloop being the second best.
Examining the best speedup over the test set, \stealing is able to obtain the best speedup (i.e., $\sim 21 \times$ on 28 threads) for the \texttt{wikipedia} sparse matrix.
On the other hand, \guided obtains its maximal speedup on 28 threads ($\sim 14.4 \times$) on the \texttt{patents} sparse matrix.
We noticed a similar trend for the worst speedup.
Overall, \dynamic and \binlpt perform very poorly.
However, \stealing  also does very poorly in the case of the sparse matrix \texttt{hugebubble}, and \ichunk can do better than \stealing in most cases. 
When comparing the sparse matrix variance (i.e., $\sigma^2$ in Table~\ref{tab:mat}) to the performance of \ichunk, a pattern starts to emerge.
For sparse matrices where variance is high (i.e., $\sigma^2 > 4.8$), \ichunk tends to do very well, i.e., it is either the best self-scheduling method or very close.
However, \ichunk does not do as well when the variance is low.
This provides some insight into why the overall geometric mean for \ichunk is not as good for \spmv as in other applications as about half (i.e., 8/15 $\sim 53\%$) of the sparse matrices have low variance, and thus, \ichunk does not perform well.
It also provides insight that \ichunk is not the best method if the user knows the workload has low variance, as the overhead in \ichunk is too high.\\

\noindent \textbf{Insight from all applications.}
Throughout these five different applications, \ichunk has performed well and been one of the top self-scheduling methods using 28 threads.
In particular, the speedup for \ichunk is never more than $\sim 10\%$ away from the best performing self-scheduling methods for the first four applications.
In the fifth application (\spmv), \ichunk is $\sim 17.5\%$ worse than the best self-scheduling method (i.e., \guided).
However, \ichunk is still in the top three best self-scheduling methods. 
Overall tested applications (including \spmv), the average speedup difference of \spmv to the best performing self-scheduling method is $\sim 5.4\%$.
As such, \ichunk does accomplish its goal of being one of the best self-scheduling methods over a wide range of irregular applications. 
However, if the application workload does not have much variance (e.g., the work per iteration is almost uniformly distributed), then \ichunk may not be the best choice.

\subsection{Sensitivity}
The previous subsection demonstrated that \ichunk could consistently provide scalable performance for a large variety of irregular applications.
However, many self-scheduling methods have some parameters such as chunk size, and \ichunk has the parameter $\epsilon$.
Some self-scheduling methods and application combinations are very sensitive to their parameters, e.g., selecting too small of a chunk size for \dynamic may result in no speedup at all.
One such example of this is \dynamic 's poor performance for the \spmv application. 
In this subsection, we go deeper into determining the sensitivity of \ichunk to the parameter $\epsilon$ and understanding how a bad choice of $\epsilon$ may result in poor performance of \ichunk.
In testing, \ichunk considers three values of $\epsilon$, which are $25\%, 33\%, \mbox{ and } 50\%$.
We define the metric: 
\begin{equation}
\epsilon\_sensitivity(app,p)  = \frac{\underset{\epsilon \in \{25\%, 33\%, 50\%\}}{max}T(app,\ichunk(\epsilon) , p)}{\underset{\epsilon \in \{25\%, 33\%, 50\%\}}{min} T(app,\ichunk(\epsilon) , p)}.
\end{equation}
This metric provides the ratio of the worst time of \ichunk over the best time of \ichunk over all three $\epsilon$ choices.
In particular, this metric will always be $>1$, and a larger value indicates a larger difference between the worst and best time.
Note that the range of $25\%$ to $50\%$ is a relatively small range when the user only considers whole number representations of $\epsilon$ compared to the chunk size range that could exist for many other self-scheduling methods.
Additionally, we note that initial experiments with $\epsilon$ outside this range generally yields bad results. 
In a more general sense, one can view $\epsilon$ as a parameter of the confidence interval around the running average of loop iterations completed as in Figure~\ref{fig:timesetp}.

\begin{figure}[t]
\centering
\includegraphics[width=1.0\textwidth]{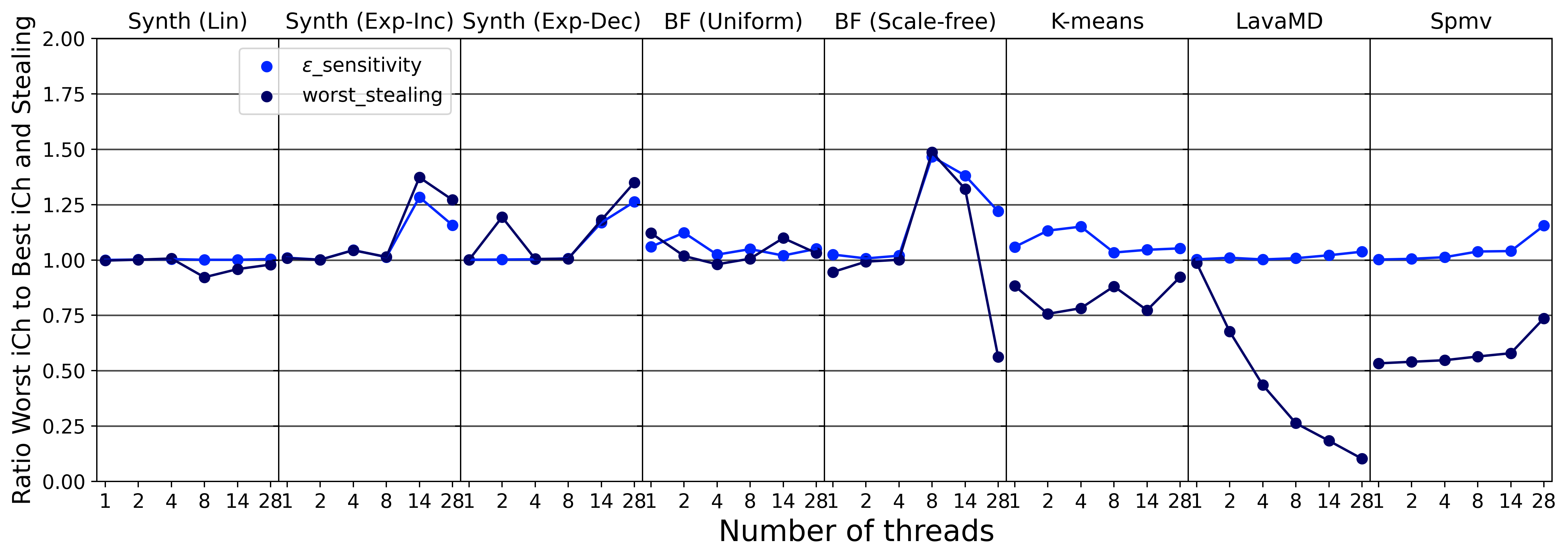}
\caption{Sensitivity.  For each application, the $\epsilon\_sensitivity$ and $worst\_stealing$ ratios are provided to help analyze the sensitivity of\ichunk to the $\epsilon$ parameter and measure how "bad" \ichunk could perform if a not using the ideal $\epsilon$. The $\epsilon\_sensitivity$ (light blue) is a ratio of the worst and best time for all tested $\epsilon$. Therefore, the larger the difference between the worst and best, the larger the ratio will be. 
The $worst\_stealing$ is the ratio of the time for \ichunk with the worst choice of $\epsilon$ over the best time for \stealing. Therefore,  a value > 1 would indicate that the best stealing method is doing better, and a value < 1 would indicate the worst \ichunk method is doing better.}
\label{fig:sense}
\end{figure}

Figure~\ref{fig:sense} presents $\epsilon\_sensitivity$ (i.e., light blue) across all the applications. 
For \synth with a linear iteration (Synth (Lin)) workload distribution, there is little to no sensitivity in the time related to the choice of $\epsilon$.
However, this is not the case for the other two distributions.
In these two cases, the right choice of $\epsilon$ could result in up to a $1.28\times$ (e.g, $\sim 28\%$) increase of time on a single socket.
Additionally, we notice that the selection of $\epsilon$ becomes more important as the thread count increases.
However, even with a poor selection of $\epsilon$, \ichunk is still able to be close to the best choice.
For example, $\epsilon\_sensitivity$ for \synth on a decreasing exponential iteration workload (Synth (Exp-Dec)) is $\sim 1.26 \times$ on 28 threads. 
However, the worst performing choice of \ichunk is still performing $\sim 2.88 \times$ faster than \guided.
Overall, the best $\epsilon$ out of the three tested for \synth for all the workloads is $25\%$.
\bfs has similar results as \synth.
That means there is little difference between the performance of \ichunk due to $\epsilon$ when the workload is somewhat evenly distributed (e.g., BF (Uniform)), and a sizable difference ($\sim 1.22\times$ on 28 threads) when the workload is far less evenly distributed.
For BF (Uniform), the best $\epsilon$ overall thread counts tends to be $25\%$.
For BF (Scale-free), the best $\epsilon$ overall thread counts tends to be $33\%$.
For \kmeans, \lava, and \spmv, we observe relatively small differences in performance, i.e., under $1.10\times$ difference.
The best $\epsilon$ for \kmeans is $33\%$, and the best $\epsilon$ for \lava is $25\%$.
The best $\epsilon$ for \spmv is highly dependent on the sparse matrix input but tends towards $33\%$.
Based on the above observations and understanding of the application, we recommend that an $\epsilon$ of $25\%$ be used on irregular applications have somewhat uniform workload distribution (and relatively small amount of work per iteration) and to increase $\epsilon$ towards $33\%$ as the workload distribution become more irregular (and relatively large amount of work per iteration). 

As a self-scheduling method built around \stealing, an important measure would be how different is the performance of \ichunk when used with a poor selection for $\epsilon$ and the performance of \stealing if we knew a good chunk size.
This type of analysis provides insight into if \ichunk is worth using over \stealing if a good parameter of $\epsilon$ is unknown.
We define the metric:
\begin{equation}
    worst\_stealing(app,p) = \frac{\underset{\epsilon \in \{25\%, 33\%, 50\%\}}{max}T(app,\ichunk(\epsilon) , p)}{\underset{chunk \in \{1,2,3,64\}}{min} T(app,\stealing(chunk) , p)}.
\end{equation}
Therefore, a value $>1$ would mean that the best \stealing method is doing better, and a value $<1$ would mean that the worst \ichunk method is doing better.
Figure~\ref{fig:sense} presents this metric (dark blue) for each of the applications.
For \synth, we observe a similar behavior to that of the previous metric.
This behavior indicates that \ichunk and \stealing are doing about the same when the loop workload is somewhat evenly distributed, but \ichunk with the poorest choice of $\epsilon$ is not as good as the best \stealing when the loop workload is not evenly distributed.
In more detail though, this behavior is not too big of a concern for \synth as \stealing does very well on this application.
For BF (Scale-free), the \stealing method with the best performance does much better than \ichunk with the worst performance on 8 and 14 threads (i.e., $\sim 1.48 \times$ and $\sim 1.32 \times$), but the case is reversed on 28 threads (i.e., $\sim .56 \times$).
Part of the reason for this turnaround is moving off the socket.  
At 28 threads, two NUMA regions exist, and failure to steal from a queue on the same socket (particularly, across the NUMA region) has a much larger penalty. 
For the other applications and inputs, \ichunk with a poor choice of $\epsilon$ does about the same or better than \stealing with the best choice of the chunk size, which is a welcomed surprise and somewhat unexpected. 
This means that \ichunk is the right choice, even if the user does not have an idea of the optimal $\epsilon$ value as for many applications it can still outperform a tuned version of \stealing.

\section{Conclusions}
Many large-scale scientific applications depend on writing code in a fork-join manner that uses parallel-for loops.
When the application is regular, i.e., the workload is evenly distributed among the iterations, a simple static scheduling that evenly distributes iterations to threads may be enough to achieve good performance. 
However, many application codes in HPC are highly irregular, i.e., the workload is not evenly distributed among iterations, and static scheduling methods do not work well.
As a result, a wide range of research has centered on constructing self-scheduling methods that help to better balance the workload among threads when assigned. 
Most of these self-scheduling methods require the user to have expert knowledge of what the workload distribution is for the particular input and to fine-tune a parameter called chunk size.
Adding to the difficulty for the user is that no self-scheduling method is good over a wide range of applications without this fine-tuning.

In this paper, we address this issue with the creation of a new self-scheduling method (called \ichunk) that aims to provide adequate performance across a wide range of applications with as little tuning as possible.
As such, this method would be one that a user could select if they do not have expert knowledge of the application and input, or they do not have time to fine-tune parameters.
The \ichunk method uses local queues to keep locally assigned loop iterations and adapts the chunk size used to assign the active chunk of iterations a thread is working on by using a cheap running approximation of iteration throughput.
When a local thread runs out of work, it steals work randomly.
The adaptive chunk size allows the thread to pace itself and tries to allow for the opportunity of other threads to steal from them without failing due to too large of an active chunk.
The \ichunk method is implemented into the OpenMP runtime system of GCC.

In evaluating, \ichunk, five different applications are used, and several of these with inputs that have very different workloads.
These applications include: Synth (\synth), Breadth-first Search (\bfs), K-Means (\kmeans), LavaMD (\lava), and Sparse Matrix-Vector Multiplication (\spmv). 
The speedup of \ichunk is compared to that of the OpenMP self-scheduling methods of \dynamic, \guided, and \taskloop.
Additionally, \ichunk is compared against the work-aware method known as \binlpt, and compared against a generic work-stealing method (\stealing) used as the base algorithm for \ichunk.
Across the various applications, \ichunk demonstrates to be the only self-scheduling method that is constantly good, as all other self-scheduling methods have at least one case where their tuned performance is worse than all the other methods.  
In particular, \ichunk is always within the top three self-scheduler methods in all applications and is always within $\sim 10\%$ of the best speedup when using 28 threads (except for \spmv).
Additionally, \ichunk is on average $\sim 5.4 \%$ away from the best speedup for all applications using 28 threads.
In analyzing the sensitivity to \ichunk's only parameter $\epsilon$, it is noted that $\epsilon$ could vary the performance by as much as $\sim 1.26\times$.
However, in most cases, the choice of $\epsilon$ is not important to achieve good performance.
Additionally, insight is provided on how to adjust $\epsilon$ if the user has some basic understanding of the workload distribution (e.g., if the distribution is somewhat uniform or very irregular).

\bibliography{omp}

\end{document}